\documentclass[letterpaper, 10 pt, conference]{ieeeconf}  % Comment this line out

\IEEEoverridecommandlockouts                              % This command is only
\usepackage{amsmath}
\usepackage{amsfonts}
\usepackage{color}
\usepackage{soul}
\usepackage{cite}
\usepackage{graphicx}
\usepackage{algorithm}
\usepackage[noend]{algpseudocode}
\usepackage{algorithmicx}
\usepackage{mathtools}
\usepackage{bm}

\usepackage{caption}
\usepackage{subcaption}
\usepackage[colorlinks=true, allcolors=blue]{hyperref}
\usepackage{adjustbox}

\DeclareMathSymbol{\shortminus}{\mathbin}{AMSa}{"39}

\usepackage{booktabs}

\DeclareMathOperator*{\argmin}{argmin}

\newcommand{\real}{\mathbb{R}}
\newcommand{\integer}{\mathbb{Z}}

\algnewcommand\algorithmicswitch{\textbf{switch}}
\algnewcommand\algorithmiccase{\textbf{case}}
\algnewcommand\algorithmicassert{\texttt{assert}}
\algnewcommand\Assert[1]{\State \algorithmicassert(#1)}%

\algdef{SE}[SWITCH]{Switch}{EndSwitch}[1]{\algorithmicswitch\ #1\ \algorithmicdo}{\algorithmicend\ \algorithmicswitch}%
\algdef{SE}[CASE]{Case}{EndCase}[1]{\algorithmiccase\ #1}{\algorithmicend\ \algorithmiccase}%
\algtext*{EndSwitch}%
\algtext*{EndCase}%

\title{\LARGE \bf
Energy-Aware Predictive Motion Planning for Autonomous Vehicles Using a Hybrid Zonotope Constraint Representation
}

\author{Joshua A. Robbins, Andrew F. Thompson, Sean Brennan, and Herschel C. Pangborn
\thanks{Joshua A. Robbins, Andrew F. Thompson, Sean Brennan, and Herschel C. Pangborn are with the Department of Mechanical Engineering, The Pennsylvania State University, University Park, PA 16802 USA (e-mail: {\tt\small jrobbins@psu.edu, thompson@psu.edu, sbrennan@psu.edu, hcpangborn@psu.edu}).}
\thanks{This research was supported by Peraton.}% <-this % stops a space
}

\begin{document}

\maketitle
\thispagestyle{empty}
\pagestyle{empty}

\begin{abstract}
Uncrewed aerial systems have tightly coupled energy and motion dynamics which must be accounted for by onboard planning algorithms. This work proposes a strategy for coupled motion and energy planning using model predictive control (MPC). A reduced-order linear time-invariant model of coupled energy and motion dynamics is presented. Constrained zonotopes are used to represent state and input constraints, and hybrid zonotopes are used to represent non-convex constraints tied to a map of the environment. The structures of these constraint representations are exploited within a mixed-integer quadratic program solver tailored to MPC motion planning problems. Results apply the proposed methodology to coupled motion and energy utilization planning problems for 1) a hybrid-electric vehicle that must 
restrict engine usage when flying over regions with noise restrictions, and 2) an electric package delivery drone that must track waysets with both position and battery state of charge requirements. By leveraging the structure-exploiting solver, the proposed mixed-integer MPC formulations can be implemented in real time.
\end{abstract}

\section{Introduction}

There has been a rising interest in the potential of autonomous electric and hybrid-electric uncrewed aerial systems (UAS) in the aviation industry. Applications include aerial package delivery vehicles~\cite{Papa2020} and air taxis or ambulances for urban air mobility (UAM)\cite{Garrow2021, nasa-uam}. Energy usage and environmental constraints, such as restrictions on aircraft noise, present significant technological challenges for these systems \cite{nasa-uam}. To address these challenges, autonomous planning algorithms must be able to account for energy utilization in addition to vehicle motion.

\subsection{Gaps in the Literature}
Existing work on energy-aware planning has focused on incorporating energy considerations into high-level path planning algorithms, often using graph-based approaches such as A* or Dijkstra's algorithm \cite{Debnath2019}. 
Graph search algorithms are used to perform high-level, energy-aware planning for hybrid-electric UAS under energy and noise constraints in \cite{Scott2022,Scott2023,Jadischke2023}, and they are applied to energy-constrained planning for package delivery drones in \cite{sorbelli2020energy, papa2020energy}. 
However, there are several key challenges when attempting to integrate high-level planners with lower-level path followers, such as discrepancies in model assumptions between the path planner and follower~\cite{Glunt2023}.

Intermediate-level motion planning algorithms are often used to bridge the gap between high-level planners and low-level controllers%in the context of motion control
. The role of such algorithms is to locally plan system trajectories, typically using a reduced-order model of the system \cite{matni2024quantitative}. Introducing energy considerations into these intermediate-level planners has received comparatively little attention in the literature despite the extensive literature on UAS motion planning when energy system dynamics are not considered \cite{Israr2022}. 

Energy and motion dynamics have been incorporated into Model Predictive Control (MPC) formulations, which are often used for motion planning \cite{gautam2024overview}. Energy dynamics were included in a hierarchical MPC controller with one spatial dimension and convex state constraints in~\cite{Koeln2018}, and a terminal battery state of charge constraint was used within an MPC path planner/follower in \cite{vallon2024learning}. In~\cite{Santos2021}, MPC is used for UAS motion planning and control. Here, battery state of charge is maximized while adhering to obstacle avoidance constraints that are imposed using potential functions, and the system dynamics are linearized about an equilibrium condition. The resulting energy-aware motion plans may be vulnerable to entrapment in suboptimal local minima given the nonlinear programming formulation and local linearization.

Increasingly, MPC optimization problems for motion planning are formulated as mixed-integer programs (MIPs) because non-convex constraint sets (e.g., an obstacle map) can be exactly represented in MIPs and---for mixed-integer convex programs---convergence to a global optimum is guaranteed \cite{ioan2021mixed}. MIPs are NP-hard \cite{floudas1995nonlinear} however, which inhibits the application of these methods in a real-time context.

\subsection{Contributions}

This paper presents an efficient method for energy-aware motion planning of uncrewed aerial systems. A mixed-integer set representation, the \emph{hybrid zonotope}, is used to exactly represent a non-convex constraint set that defines obstacles and regions with location-specific noise restrictions. A reduced-order, linear time-invariant model of the coupled UAS energy and motion dynamics is developed that is globally valid and conservative with respect to planned energy usage. Motion and energy states are coupled via a polytopic constraint set, and a low-complexity \emph{constrained zonotope} representation of this set is presented. An MPC controller is formulated that plans system trajectories which adhere to specifications on both the motion and energy states. A mixed-integer quadratic program (MIQP) solver developed in our previous work \cite{robbins2024efficient, robbins2024efficientjournal} (previously applied for motion planning only and not energy management) is leveraged to efficiently solve these MPC optimization problems by exploiting the structure of the hybrid zonotope and constrained zonotope set representations. Case studies show how noise-restricted areas and terminal energy constraints can be considered by the proposed controller and highlight the utility of jointly optimizing energy and motion plans.

\section{Preliminaries} \label{sec:prelims}

\subsection{Notation}
Vectors are denoted with boldface letters. Sets are denoted with calligraphic letters. Vertex representation (V-rep) polytopes are denoted in terms of their vertices $\mathbf{v}_i$ as $\mathcal{P} = \{\mathbf{v}_1, \mathbf{v}_2, ...\}$. Empty brackets $[\;]$ denote the absence of a quantity. Expressions using the $\pm$ symbol are expanded using all possible permuatations. For instance, $\pm a \pm b \leq c$ expands to the inequalities
\begin{equation}
\begin{matrix}
    a + b \leq c\,, & -a + b \leq c\,,\\
    a - b \leq c\,, & -a - b \leq c\,.       
\end{matrix}
\end{equation}

\subsection{Zonotopes, Constrained Zonotopes, and Hybrid Zonotopes} \label{sec:hybzono-definition}

As will be seen in Sec.~\ref{sec:formulation}, constrained zonotopes and hybrid zonotopes are used to represent constraint sets in our MPC formulation. The definitions of these set representations are briefly reviewed here.

A set $\mathcal{Z} \subset \real^n$ is a zonotope if $\exists \; G_c \in \real^{n \times n_g}$, $\mathbf{c} \in \real^{n}$ such that
\begin{equation} \label{eq:zonotope}
\mathcal{Z} = \left\{ G_c \bm{\xi}_c + \mathbf{c} \middle| \bm{\xi}_c \in \mathcal{B}_{\infty}^{n_g} \right\} \;,
\end{equation}
where $\mathcal{B}_\infty^{n_g} = \{\xi_c\in\mathbb{R}^{n_g} \mid \lVert \xi_c \rVert_\infty\leq1\}$ is the infinity-norm ball. Zonotopes are convex, centrally symmetric sets \cite{ziegler2012lectures}.

A set $\mathcal{Z_C} \subset \real^n$ is a constrained zonotope if $\exists \; G_c \in \real^{n \times n_g}$, $\mathbf{c} \in \real^{n}$, $A_c \in \real^{n_c \times n_g}$, $\mathbf{b} \in \real^{n_c}$ such that
\begin{equation} \label{eq:cons_zonotope}
\mathcal{Z_C} = \left\{ G_c \bm{\xi}_c + \mathbf{c} \middle| \bm{\xi}_c \in \mathcal{B}_{\infty}^{n_g},\; A_c \bm{\xi}_c = \mathbf{b} \right\} \;.
\end{equation} 
Constrained zonotopes can represent any polytope \cite{scott2016constrained}.

Hybrid zonotopes extend \eqref{eq:cons_zonotope} by including binary factors $\bm{\xi}_b$. A set $\mathcal{Z_H} \subset \real^n$ is a hybrid zonotope if in addition to $G_c$, $\mathbf{c}$, $A_c$, and $\mathbf{b}$, $\exists\; G_b \in \real^{n \times n_b}$, $A_b \in \real^{n_c \times n_b}$ such that
\begin{equation} \label{eq:hyb_zonotope}
\mathcal{Z_H} = \left\{ \begin{bmatrix} G_c & G_b \end{bmatrix} 
\begin{bmatrix} \bm{\xi}_c \\ \bm{\xi}_b \end{bmatrix} + \mathbf{c} \middle| 
\begin{matrix}
\begin{bmatrix} \bm{\xi}_c \\ \bm{\xi}_b \end{bmatrix} \in \mathcal{B}_{\infty}^{n_g} \times \{-1,1\}^{n_b} \\
\begin{bmatrix} A_c & A_b \end{bmatrix} \begin{bmatrix} \bm{\xi}_c \\ \bm{\xi}_b \end{bmatrix} = \mathbf{b}
\end{matrix}
\right\} \;.
\end{equation}
Hybrid zonotopes can represent any union of polytopes \cite{bird2023hybrid}.

In this paper, hybrid zonotopes are denoted using the shorthand notation $\mathcal{Z}_H = \left\langle G_c, G_b, \mathbf{c}, A_c, A_b, \mathbf{b} \right\rangle$. Analogously, constrained zonotopes are denoted with $\mathcal{Z}_C = \left\langle G_c, \mathbf{c}, A_c, \mathbf{b} \right\rangle$.

\section{Prediction Model Formulation} \label{sec:formulation}

We consider a UAS that must navigate to a reference position through a non-convex feasible space (e.g., a map containing obstacles). The UAS velocity is constrained by the output power of onboard energy systems. In one numerical example (\emph{Case Study 1} in Sec.~\ref{sec:results}), we address a fixed-wing UAS with a hybrid-electric powertrain where the environment includes areas with noise restrictions such that the engine power is limited. In another example (\emph{Case Study 2} in Sec.~\ref{sec:results}), we address an electric package delivery drone which must navigate to a wayset that includes constraints on the battery state of charge.

In this section, a linear time-invariant, reduced-order model of these systems is presented for use in a predictive controller. Differential flatness is used to account for non-linearities in the motion dynamics. The linear energy dynamics are coupled to the motion dynamics via a polytopic state constraint set (represented as a constrained zonotope) in such a way that the planned energy utilization is conservative under a quasi-steady assumption. Non-convex constraints are represented exactly as a hybrid zonotope.

\subsection{Reduced-Order Model} \label{sec:reduced-order-model}
A linear time-invariant, reduced-order model of the coupled UAS motion and energy dynamics is presented here. The dynamics of the motion states are given by the unicycle model
\begin{equation} \label{eq:unicycle}
    \dot{\xi} =  v \cos{\theta} \;,\; \dot{\eta} = v \sin{\theta} \;,\; \dot{\theta} = \omega \;,\;
\end{equation}
with $\xi$ and $\eta$ the vehicle position, $\theta$ the heading angle, $v$ the velocity, and $\omega$ the turn rate. The unicycle model is a commonly used reduced-order model in many domains and has been applied to UAS \cite{panyakeow2010decentralized}. This model is differentially flat with respect to the flat outputs $\xi$ and $\eta$, meaning that the states can be expressed in terms of the flat outputs and their derivatives \cite{diffflat}. Differential flatness is frequently exploited when possible to account for non-linearities in motion planning \cite{ioan2021mixed, agrawal2021constructive, matni2024quantitative}. For the unicycle model, the states are expressed in terms of the flat outputs as
\begin{equation}
    v = \sqrt{\dot{\xi}^2 + \dot{\eta}^2} \;,\; \theta = \text{atan2}(\dot{\eta},\; \dot{\xi}) \;,\; \omega = \frac{\dot{\xi} \ddot{\eta} - \dot{\eta} \ddot{\xi}}{\dot{\xi}^2 + \dot{\eta}^2} \;.
\end{equation}

The unicycle dynamics in terms of the flat outputs can then be expressed by the double integrator model
\begin{equation} \label{eq:dbl-int-dyn}
    \begin{bmatrix}
        \dot{\xi} \\ \ddot{\xi} \\ \dot{\eta} \\ \ddot{\eta}
    \end{bmatrix} = 
    \begin{bmatrix}
        0 & 1 & 0 & 0 \\
        0 & 0 & 0 & 0 \\
        0 & 0 & 0 & 1 \\
        0 & 0 & 0 & 0
    \end{bmatrix} \begin{bmatrix}
        \xi \\ \dot{\xi} \\ \eta \\ \dot{\eta}
    \end{bmatrix} +
    \begin{bmatrix}
        0 & 0 \\
        1 & 0 \\
        0 & 0 \\
        0 & 1
    \end{bmatrix} \begin{bmatrix}
        \ddot{\xi} \\ \ddot{\eta}
    \end{bmatrix} \;.
\end{equation}
In contrast with \eqref{eq:unicycle}, \eqref{eq:dbl-int-dyn} is a linear time-invariant (LTI) model. LTI dynamics are required for the MPC formulation given in~\eqref{eq:mpc-lin-dyn}. A velocity limit $v_{lim}$ and turn rate limit $\omega_{lim}$ are conservatively enforced using the polytopic constraints \cite{whitaker2021optimal}
\begin{subequations} \label{eq:motion-constr}
\begin{align}
    &\pm \dot{\xi} \pm \dot{\eta} \leq v_{lim} \;, \label{eq:motion-constr-vel} \\
    &\pm \ddot{\xi} \pm \ddot{\eta} \leq v_{min} \omega_{lim} \;, \label{eq:motion-constr-omega}
\end{align}
\end{subequations}
where $v_{min}$ is the minimum velocity of the vehicle. For the case that a minimum velocity must be strictly enforced (e.g., for a fixed-wing aircraft as in \emph{Case Study 1}), the constraint 
\begin{equation} \label{eq:xidot-geq-vmin}
\dot{\xi} \geq v_{min} \;,
\end{equation}
can be added to the state constraints. Applying this constraint imposes a requirement for forward progress along the $\xi$ direction. This is most applicable for vehicles with limited maneuverability over the planning horizon or for the case that $\xi$ and $\eta$ are defined as path-relative coordinates given a global path plan. Alternatively, a less restrictive constraint such as $\pm \dot{\xi} \pm \dot{\eta} \geq v_{min}$ could be used at the expense of the state constraint set becoming non-convex.

For the case of a hybrid-electric vehicle (\emph{Case Study 1}), the energy dynamics are given by the simplified first-order model
\begin{multline} \label{eq:energy-dyn}
    \begin{bmatrix}
        \dot{SOC} \\ \dot{P}_b \\ \dot{m}_f \\ \dot{P}_e
    \end{bmatrix} = 
    \begin{bmatrix}
        0 & -1/C_b & 0 & 0 \\
        0 & 0 & 0 & 0 \\
        0 & 0 & 0 & -SFC \\
        0 & 0 & 0 & 0
    \end{bmatrix}
    \begin{bmatrix}
        SOC \\ P_b \\ m_f \\ P_e
    \end{bmatrix} + \\
    \begin{bmatrix}
        0 & 0 \\
        1 & 0 \\
        0 & 0 \\
        0 & 1
    \end{bmatrix}
    \begin{bmatrix}
        \dot{P}_b \\ \dot{P}_e
    \end{bmatrix} \;,
\end{multline}
where $SOC$ is the battery state of charge, $C_b$ is the battery capacity, $m_f$ is the fuel mass, and $SFC$ is the specific fuel consumption. $P_b$ and $P_e$ are the battery and engine power outputs, respectively. All states and inputs in \eqref{eq:energy-dyn} are subject to box constraints. Negative battery powers are permitted (i.e.,  $P_{b min}<0$) to allow for battery charging. A minimum total output power is also enforced such that
\begin{equation} \label{eq:P_min}
    P_b + P_e \geq P_{min} \;,
\end{equation}
where $P_{min}$ corresponds to the power needed to maintain a minimum velocity $v_{min}$. For \emph{Case Study 2}, there is no engine so all states, inputs, and constraints associated with the engine are eliminated.

\begin{figure}[tb]
    \centering
    \includegraphics[width=0.8\linewidth]{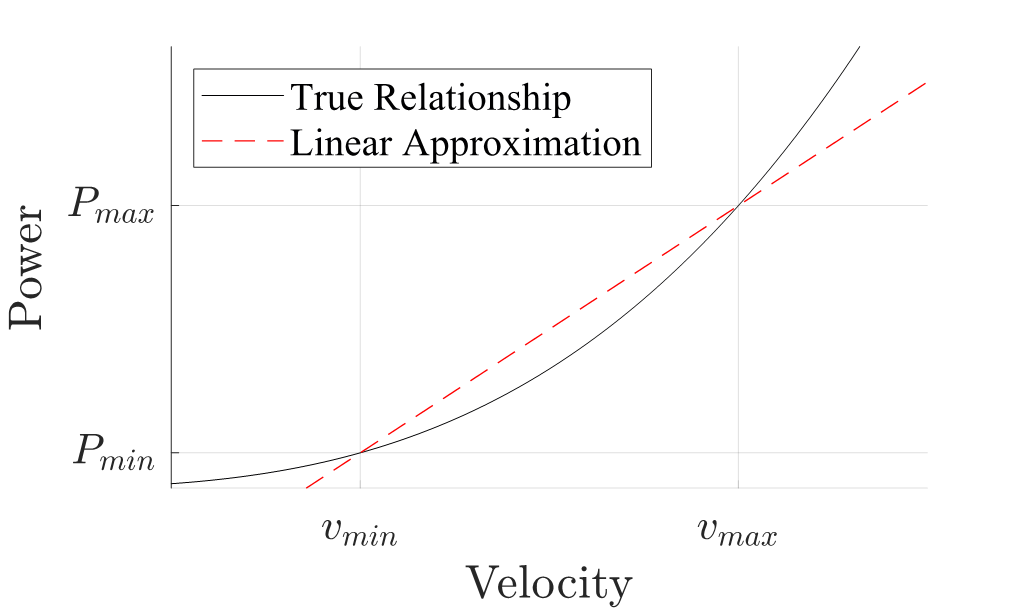}
    \caption{Linear, quasi-steady approximation of the velocity and power relationship.}
    \label{fig:power-vel}
\end{figure}

The motion and energy dynamics are coupled using a linear, quasi-steady approximation of the relationship between output power $P$ and velocity $v$ as shown in Fig.~\ref{fig:power-vel}. For aerial vehicles, typically the relationship $P \propto v^3$ approximately holds in steady state. This follows from $F_D \propto v^2$ where $F_D$ is the drag force magnitude \cite{hoerner}. A linear approximation of this relationship is
\begin{equation} \label{eq:lin_P_v}
    P = \frac{P_{max}-P_{min}}{v_{max}-v_{min}} (v - v_{min}) + P_{min} \;.
\end{equation}
This particular linear approximation is chosen such that the planned power usage from the energy-aware motion planner will always be greater than the power usage for a nonlinear quasi-steady model. 

The velocity limit $v_{lim}$ in \eqref{eq:motion-constr-vel} is required to be less than or equal to the quasi-steady velocity $v$ given power $P$ in \eqref{eq:lin_P_v}. Substituting \eqref{eq:lin_P_v} into \eqref{eq:motion-constr-vel} gives 
\begin{multline} \label{eq:state-constr-hrep}
    \pm \dot{\xi} \pm \dot{\eta} - \left( \frac{v_{max}-v_{min}}{P_{max}-P_{min}} \right) (P_b + P_e) \leq \\ v_{min} -\left( \frac{v_{max}-v_{min}}{P_{max}-P_{min}} \right) P_{min} \;,
\end{multline}
which couples the motion dynamics to the energy dynamics.

The state and input vectors for the coupled system are then given as
\begin{subequations}
\begin{align}
    &\mathbf{x} = \begin{bmatrix} \xi & \dot{\xi} & \eta & \dot{\eta} & SOC & P_b & m_f & P_e \end{bmatrix}^T \;, \\
    &\mathbf{u} = \begin{bmatrix} \ddot{\xi} & \ddot{\eta} & \dot{P}_b & \dot{P}_e \end{bmatrix}^T \;.
\end{align}
\end{subequations}

\subsection{Constrained Zonotope State and Input Constraints} \label{sec:conzono-constraints}

\begin{figure}[tb]
    \centering
        \includegraphics[width=0.7\linewidth]{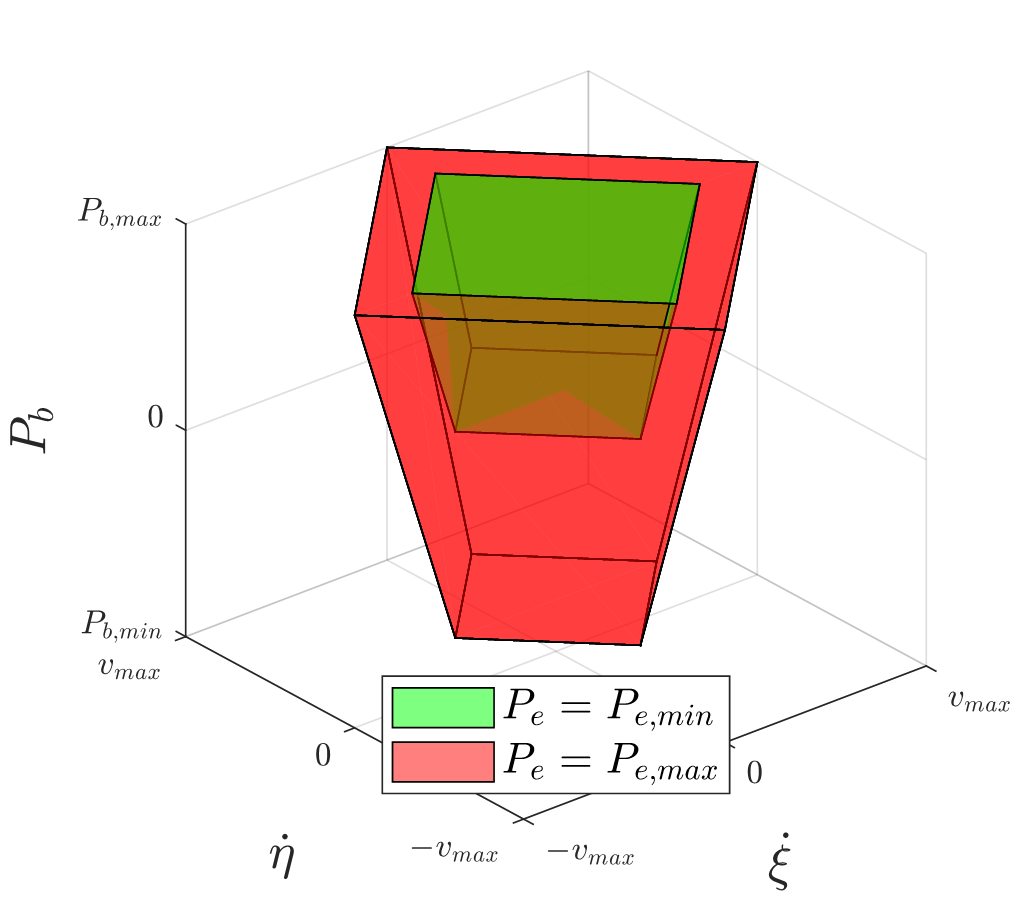}
        
        \vspace{-0.05in}
        
        \includegraphics[width=0.7\linewidth]{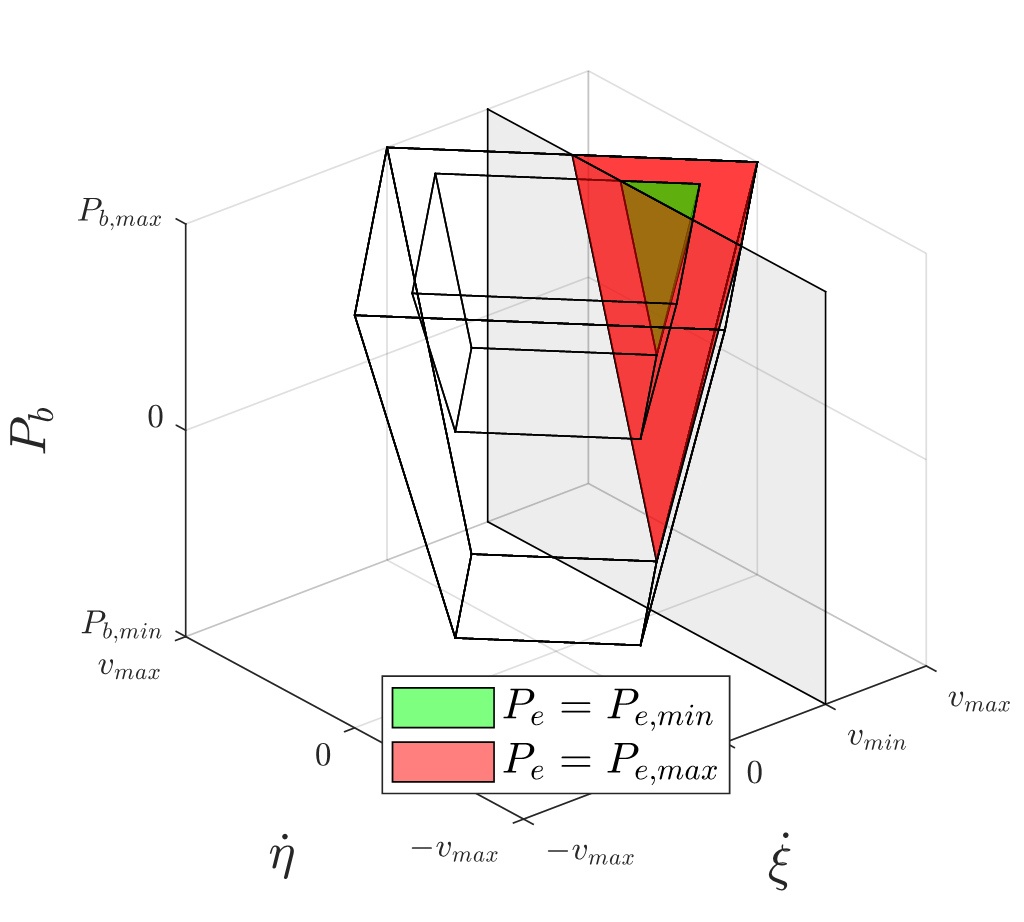}
    \caption{Projections of the constrained zonotope $\mathcal{Z}_{cx}$ representing coupled constraints on the energy and motion states. The bottom sub-figure shows the case where the forward progress constraint \eqref{eq:xidot-geq-vmin} is imposed.}
    \label{fig:state-const-conzono}
\end{figure}

As will be discussed in Sec.~\ref{sec:miqp-solver}, the energy-aware motion planning problem can be efficiently solved in part by exploiting a constrained zonotope representation of the state and input constraints. Constrained zonotopes with fewer factors ($\bm{\xi}_c$ in \eqref{eq:cons_zonotope}) or equality constraints facilitate more efficient optimization. Equality constraints in particular should be minimized, as discussed in \cite[Sec.~III-C.2]{robbins2024efficient}. 

The polytopic constraint for the inputs $\ddot{\xi}$ and $\ddot{\eta}$ is given in halfspace representation (H-rep) in \eqref{eq:motion-constr-omega}. This is equivalently expressed as the constrained zonotope
\begin{equation} \label{eq:input-conzono}
    \mathcal{Z}_{cu} = \left\langle \frac{1}{2} v_{min} \omega_{max} \begin{bmatrix}
        1 & 1 \\ -1 & 1
    \end{bmatrix}, \begin{bmatrix}
        0 \\ 0
    \end{bmatrix}, [], [] \right\rangle \;,
\end{equation}
such that $\begin{bmatrix} \ddot{\xi}^T & \ddot{\eta}^T \end{bmatrix}^T \in \mathcal{Z}_{cu}$. Note that \eqref{eq:input-conzono} can be interpreted as a rotated box constraint. 

The state variables $\dot{\xi}$, $\dot{\eta}$, $P_b$ and $P_e$ are subject to the polytopic constraint defined in H-rep by \eqref{eq:P_min} and \eqref{eq:state-constr-hrep} in addition to the box constraints on $P_b$ and $P_e$. This polytopic constraint is equivalently written as the constrained zonotope
\begin{multline} \label{eq:state-conzono}
    \mathcal{Z}_{cx} = \left\langle \begin{bmatrix}
        0 & 0 & \frac{b_z}{2} & -\frac{b_z}{2} & \frac{b_z}{2} & -\frac{b_z}{2} & 0 \\
        0 & 0 & -\frac{b_z}{2} & \frac{b_z}{2} & \frac{b_z}{2} & -\frac{b_z}{2} & 0 \\
        g_b & 0 & 0 & 0 & 0 & 0 & 0 \\
        0 & g_e & 0 & 0 & 0 & 0 & 0
    \end{bmatrix}, \begin{bmatrix}
        0 \\ 0 \\ c_b \\ c_e
    \end{bmatrix},  \right. \\ \left. \begin{bmatrix}
        g_b & g_e & a_z & a_z & 0 & 0 & 0 \\
        0 & 0 & a_z & a_z & 0 & 0 & c_z \\
        0 & 0 & 0 & 0 & a_z & a_z & c_z
    \end{bmatrix}, \begin{bmatrix}
        c_1 - c_b - c_e \\ a_z \\ a_z
    \end{bmatrix} \right\rangle \;,
\end{multline}
where $\begin{bmatrix} \dot{\xi} & \dot{\eta} & P_b & P_e \end{bmatrix}^T \in \mathcal{Z}_{cx}$ and
\begin{subequations} \label{eq:state-conzono-vars}
\begin{align}
    &c_z = \frac{P_{max} - P_{min}}{2} \;, \\
    &a_z = \left( \frac{v_{max} + v_{min}}{v_{max} - v_{min}} \right) c_z \;, \\
    &b_z = \left( \frac{a_z}{a_z + c_z} \right) v_{max} \;, \\
    &g_b = \frac{P_{b max} - P_{b min}}{2} \;, \\
    &c_b = \frac{P_{b max} + P_{b min}}{2} \;, \\ 
    &g_e = \frac{P_{e max} - P_{e min}}{2} \;, \\
    &c_e = \frac{P_{e max} + P_{e min}}{2} \;, \\
    &c_1 = 2 a_z - (a_z-c_z) + P_{min} \;.
\end{align}
\end{subequations}
This formulation of $\mathcal{Z}_{cx}$ has 7 factors and 4 equality constraints. For comparison, the standard method of constructing $\mathcal{Z}_{cx}$ would be use an H-rep polytope to constrained zonotope conversion \cite[Thm 1]{scott2016constrained}. This would result in a constrained zonotope with 13 factors and 9 equality constraints. At the cost of an additional factor and equality constraint, \eqref{eq:xidot-geq-vmin} can be incorporated into to $\mathcal{Z}_{cx}$ using the identity for the intersection of a constrained zonotope and a single halfspace inequality given in \cite{raghuraman2022set}.

Fig.~\ref{fig:state-const-conzono} depicts projections of $\mathcal{Z}_{cx}$ onto the $\dot{\xi}$, $\dot{\eta}$, and $P_b$ states for $P_e$ held at its maximum and minimum values. This figure shows how energy and motion states are coupled within the reduced-order model.

\subsection{Hybrid Zonotope Output Constraints}
Non-convex constraints are used in this model to describe obstacle avoidance constraints or keep out areas. Additionally, they are used to describe areas where there are restrictions on the noise generated by the vehicle, as in \emph{Case Study 1}.
To represent the non-convex constraint set $\mathcal{F} \subset \real^n$ as a hybrid zonotope, a vertex representation to hybrid zonotope conversion \cite[Thm 5]{siefert2023reachability} is used as implemented in \cite{koeln2023zonolab}. As shown in \cite{robbins2024efficientjournal}, hybrid zonotopes constructed this way have the property that their convex relaxation is their convex hull. This property is exploited when solving the energy-aware motion planning problems using the solver described in Sec.~\ref{sec:miqp-solver}.

To construct the vertex representation of $\mathcal{F}$, the free space around any obstacles and noise-restricted areas is first partitioned using the Hertel and Mehlhorn algorithm \cite{o1998computational}. Polytope $i$ in the free-space partition is defined in terms of its vertices as $\mathcal{P}^f_i = \{\mathbf{v}^f_{i 1}, \cdots, \mathbf{v}^f_{i n_i} \}$, and the noise-restricted areas $\mathbf{P}^r_i$ are similarly defined. For the case of a hybrid-electric vehicle, noise is assumed, as in~\cite{Scott2022,Scott2023,Jadischke2023}, to be generated primarily by the engine such that noise restriction constraints are achieved for $P_e \leq P_{noise}$. The noise restriction constraints are then captured by extending the $\mathcal{P}^f_i$ and $\mathcal{P}^r_i$ to a third dimension such that 
\begin{subequations}
\begin{align}
    &\mathcal{F} = \left( \bigcup_i \mathcal{F}^f_i \right) \cup \left( \bigcup_j \mathcal{F}^r_j \right) \;, \\
    &\mathcal{F}^f_i = \left\{ \begin{bmatrix} \mathbf{v}^f_{i 1} \\ 0 \end{bmatrix}, \cdots, \begin{bmatrix} \mathbf{v}^f_{i n_i} \\ 0 \end{bmatrix}, \begin{bmatrix} \mathbf{v}^f_{i 1} \\ P_{e max} \end{bmatrix}, \cdots, \begin{bmatrix} \mathbf{v}^f_{i n_i} \\ P_{e max} \end{bmatrix} \right\} \;, \\
    &\mathcal{F}^r_j = \left\{ \begin{bmatrix} \mathbf{v}^r_{j 1} \\ 0 \end{bmatrix}, \cdots, \begin{bmatrix} \mathbf{v}^r_{j n_j} \\ 0 \end{bmatrix}, \begin{bmatrix} \mathbf{v}^r_{j 1} \\ P_{noise} \end{bmatrix}, \cdots, \begin{bmatrix} \mathbf{v}^r_{j n_j} \\ P_{noise} \end{bmatrix} \right\} \;.
\end{align}
\end{subequations}
\section{Controller Formulation} \label{sec:controller-formulation}
In this section, a model predictive control formulation and associated solution methodology are presented that efficiently perform energy-aware motion planning using the reduced-order UAS model developed in Sec.~\ref{sec:formulation}.

\subsection{MPC Formulation} \label{sec:mpc-formulation}
Consider the following MPC formulation:
\begin{subequations} \label{eq:mpc-gen}
\begin{align}
    &\min_{\mathbf{x}_k, \mathbf{u_k}} \sum_{k=0}^{N-1} \left[ (\mathbf{x}_k - \mathbf{x}_k^r)^T Q_k (\mathbf{x}_k - \mathbf{x}_k^r) + \mathbf{u}_k^T R_k \mathbf{u}_k + \right. \nonumber \\
    & \qquad \left. \mathbf{q}_k^T \mathbf{x}_k + q^r(\mathbf{y}_k) \right] + (\mathbf{x}_N - \mathbf{x}_N^r)^T Q_N (\mathbf{x}_N - \mathbf{x}_N^r) + \nonumber \\
    & \qquad \left. \mathbf{q}_N^T \mathbf{x}_N + q^r(\mathbf{y}_N)\;, \right.\\
    &\text{s.t.} \; \forall k \in \mathcal{K} = \{0, \cdots, N-1 \}\;: \nonumber \\ 
    &\phantom{\text{s.t.}} \; \mathbf{x}_{k+1} = A \mathbf{x}_k + B \mathbf{u}_k\;, \label{eq:mpc-lin-dyn}\\
    &\phantom{\text{s.t.}} \; \mathbf{y}_{k} = H \mathbf{x}_k,\; \mathbf{y}_{N} = H \mathbf{x}_N,\; \mathbf{x}[0] = \mathbf{x}_0 \;, \\
    &\phantom{\text{s.t.}} \; \mathbf{x}_k, \mathbf{x}_k^r \in \mathcal{X},\; \mathbf{x}_N, \mathbf{x}_N^r \in \mathcal{X_T},\;\mathbf{u}_k \in \mathcal{U}\;, \label{eq:mpc-gen-state-input-cons} \\
    &\phantom{\text{s.t.}} \; \mathbf{y}_{k}, \mathbf{y}_{N} \in \mathcal{F} = \bigcup_{i=1}^{n_F} \mathcal{F}_i \subset \real^n \;, \label{eq:mpc-gen-obs-avoid} \\
    &\phantom{\text{s.t.}} \; \mathbf{y} \in \mathcal{F}_i \Rightarrow q^r(\mathbf{y}) = q_{i}^r
    \;. \label{eq:mpc-gen-reg-dep-costs}
\end{align}
\end{subequations}
The state and input are given by $\mathbf{x}_k$ and $\mathbf{u}_k$, respectively. A state reference to be tracked is given by $\mathbf{x}_k^r$, and the MPC horizon is $N$. The initial state of the system is $\mathbf{x}[0]$. The sets $\mathcal{X}$, $\mathcal{X}_N$, $\mathcal{U}$, and $\mathcal{F}_i \; \forall i \in \{1, ..., n_F\}$ are assumed be convex polytopes. The outputs $\mathbf{y}_k$, $\mathbf{y}_N$ are constrained to a set $\mathcal{F}$ which is the union of polytopes $\mathcal{F}_i$. The function $q^r: \real^n \rightarrow \real$ couples the system outputs to region dependent costs via \eqref{eq:mpc-gen-reg-dep-costs}.

\subsection{MPC Cost Function}
Power states $P_b$ and $P_e$ are related to velocity states $\dot{\xi}$ and $\dot{\eta}$ via the inequality constraint \eqref{eq:state-constr-hrep}. This constraint is an inner approximation of $\sqrt{\dot{\xi}^2 + \dot{\eta}^2} \leq v_{lim}$ where $v_{lim}$ is a function of the total output power $P_b + P_e$. The planned power usage will in general be greater than is required to achieve a given velocity $v = \sqrt{\dot{\xi}^2 + \dot{\eta}^2}$. In order to reduce this discrepancy, a linear cost is placed on $P_b + P_e$ via the $\mathbf{q}_k$ and $\mathbf{q}_N$ terms in \eqref{eq:mpc-gen}.

\subsection{Efficient Solution of Multi-State MIQPs with Constrained Zonotope and Hybrid Zonotope Constraint Representations} \label{sec:miqp-solver}
The MPC formulation given in \eqref{eq:mpc-gen} can be rewritten as the following multi-stage mixed integer quadratic program (MIQP):
\begin{subequations} \label{eq:miqp-multistage}
\begin{align} 
&\mathbf{z}^* = \argmin_{\mathbf{z}} \sum_{k=0}^{N} \frac{1}{2} \mathbf{z}_k^T P_k \mathbf{z}_k + \mathbf{q}_k^T \mathbf{z}_k\;, \label{eq:miqp-multistage-cost} \\
&\text{s.t.} \; \mathbf{0} = C_k \mathbf{z}_k + D_{k+1} \mathbf{z}_{k+1} + \mathbf{c}_k,\; \forall k \in \mathcal{K}\;, \label{eq:miqp-multistage-eqcons}\\
&\phantom{\text{s.t.}} \; G_k \mathbf{z}_k \leq \mathbf{w}_k,\; \forall k \in \mathcal{K} \cup N \;,\label{eq:miqp-multistage-ineqcons}
\end{align}
\end{subequations}
with $\mathbf{z}_k = \begin{bmatrix} \mathbf{x}_k^T & \mathbf{u}_k^T & \bm{\alpha}_{k}^T \end{bmatrix}^T$. The vector $\bm{\alpha}_k \in \real^{n_{ck}} \times \integer^{n_{bk}}$ denotes additional continuous and integer variables used to define constraints and slack variables. 

Problem \eqref{eq:miqp-multistage} is formulated using constrained zonotope representations of $\mathcal{X}$, $\mathcal{X}_N$, $\mathcal{U}$ and a hybrid zonotope representation of $\mathcal{F}$ in \cite{robbins2024efficientjournal}. The structure of these constraint representations is exploited when solving \eqref{eq:miqp-multistage} in the multi-stage MIQP solver presented in \cite{ robbins2024efficientjournal}. This solver uses these set representations to reduce the number of matrix factorizations that need to be performed in quadratic program sub-problems and to reduce the number of iterations required to converge in a branch-and-bound mixed-integer solver. 
The structure-exploiting MIQP solver is written in C++ and called from MATLAB. 

For comparison, equivalent formulations of the MIQPs are constructed using more traditional set representations. Specifically, H-rep polytopes are used instead of constrained zonotopes and unions of H-rep polytopes via the Big-M method \cite{ioan2021mixed} are used instead of hybrid zonotopes. These equivalent MIQPs are solved with the state-of-the-art commercial solver Gurobi via its MATLAB API \cite{gurobi}.
\section{Results} \label{sec:results}
This section evaluates the proposed energy-aware motion planner in two different case studies. \emph{Case Study 1} concerns a fixed-wing hybrid-electric UAS model. \emph{Case Study 2} concerns an electric package delivery drone. In both cases, the planner must navigate the vehicle to a terminal reference state $\mathbf{x}^r$. An Ubuntu 22.04 desktop with an i7-14700 processor and 32GB of RAM is used to solve the MIQPs, and the solvers are configured to use up to 16 threads. Absolute and relative convergence tolerances are set to $\epsilon_a = 0.1$ and $\epsilon_r = 0.01$ respectively, and a 15-step MPC prediction horizon is used.

\subsection{Case Study 1: Hybrid-Electric UAS with Noise-Restricted Areas}

\begin{figure}
    \centering
    \includegraphics[width=0.6\linewidth]{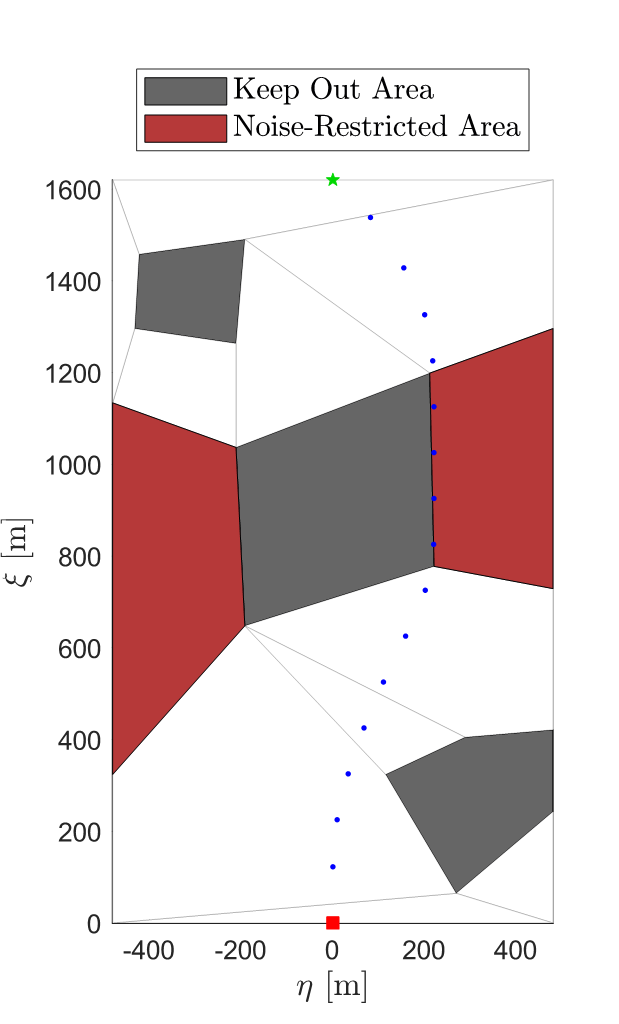}
    \caption{\emph{Case Study 1:} Planned trajectory for a hybrid-electric UAS. The red square is the start position and the green star corresponds to the reference state $\mathbf{x}_N^r = \mathbf{x}_k^r \; \forall k \in \{0, ..., N-1\}$. The blue dots are the planned vehicle trajectory.}
    \label{fig:ex1-traj}
\end{figure}

\begin{figure}
    \centering
    \includegraphics[width=\linewidth]{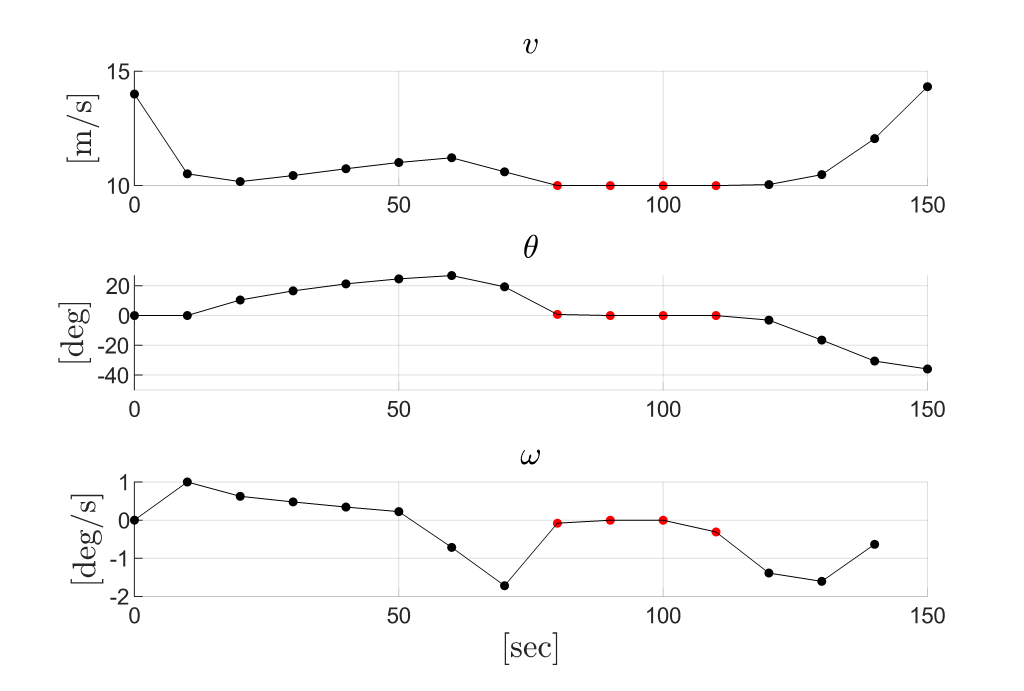}
    \caption{\emph{Case Study 1:} Planned motion states and inputs for the hybrid-electric UAS example. Red points indicate time steps where the vehicle is in the noise-restricted area.}
    \label{fig:ex1-motion-states}
\end{figure}

\begin{figure}
    \centering
    \includegraphics[width=\linewidth]{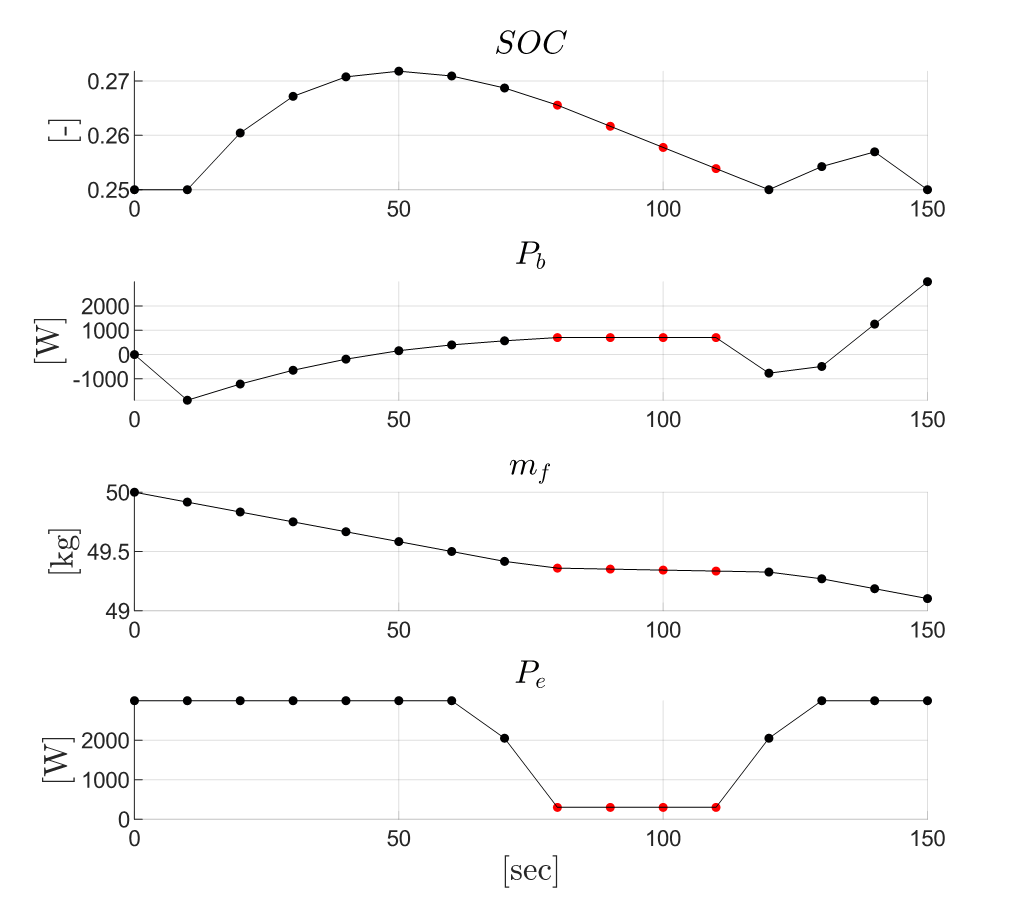}
    \caption{\emph{Case Study 1:} Planned energy states for the hybrid-electric UAS example. Red points indicate time steps where the vehicle is in the noise-restricted area.}
    \label{fig:ex1-energy-states}
\end{figure}

In this example, a fixed wing hybrid-electric UAS is considered. The vehicle must navigate through a map with both keep-out areas and areas with noise restrictions. In the noise-restricted areas, the engine power is limited to $P_e \in [P_{e min}, P_{noise}]$.

A discretized version of the continuous-time dynamics model given in Sec.~\ref{sec:reduced-order-model} is used with a discrete time step of $\Delta t = 10$~sec. The battery capacity and specific fuel consumption are $C_b = 0.5$~kWh and $SFC = 10$~kg/kWh, respectively. Limits for the states are given in Table~\ref{tab:ex1-state-limits}. The values used for these parameters (and those in the second case study) are not intended to be representative of any specific vehicle, but are instead chosen to highlight the features of the proposed approach. The coupled state constraints \eqref{eq:state-conzono} are modified to include the forward progress constraint \eqref{eq:xidot-geq-vmin} as described in Sec.~\ref{sec:conzono-constraints}. The terminal state constraint set is set to $\mathcal{
X}_N = \mathcal{X}$. The engine power limit within the noise-restricted areas is $P_{noise} = 300$~W. 

\begin{table}[htbp]
  \setlength{\tabcolsep}{2pt}
    \centering
        \caption{\emph{Case Study 1:} State and input limits for hybrid-electric aircraft model}
    \begin{tabular}{c|c c c c c c c c c}
         \textbf{Parameter} & $v$ & $\omega$ & $SOC$ & $m_f$ & $P$ & $P_b$ & $P_e$ & $\dot{P}_b$ & $\dot{P}_e$ \\ \toprule
         \textbf{units} & [m/s] & [deg/s] & [-] & [kg] & [kW] & [kW] & [kW] & [W/s] & [W/s] \\ \midrule
         \textbf{min} & 10 & -2 & 0.25 & 0 &  1.0 & -3.0 & 0.0 & -1400 & -175 \\
         \textbf{max} & 20 & 2 & 1 & 50 & 6.0 & 3.0 & 3.0 & 1400 & 175
    \end{tabular}
    \label{tab:ex1-state-limits}
\end{table}

Referencing \eqref{eq:mpc-gen}, the cost function parameters used in this example are
\begin{subequations}
\begin{align}
    &Q_k = 0_{8 \times 8} \;, \\
    &Q_N = \text{diag}([10^{-2}, 0, 10^{-2}, 0, 0, 0, 0, 0]) \;, \\
    &R_k = \text{diag}([1, 1, 10^{-4}, 10^{-4}]) \;, \\
    &\mathbf{q}_k = \mathbf{q}_N = [0, 0, 0, 0, 0, 10^{-3}, 0, 10^{-3}]^T \;, \\
    &q^r(\mathbf{x}_k) = 0 \; \forall \mathbf{x}_k \;.
\end{align}
\end{subequations}

The computed motion and energy utilization plan for this example is given in Figs.~\ref{fig:ex1-traj} to \ref{fig:ex1-energy-states}. The system starts with the battery at its minimum allowable state of charge and so does not have sufficient onboard energy to pass through the noise-restricted areas. Accordingly, the energy-aware motion planner selects a position and velocity profile for which it can charge the battery enough to power the vehicle through the noise-restricted area prior to entering that area. The energy system trajectories are jointly optimized to provide sufficient power for the planned velocity profile. The engine power adheres to noise restriction specifications but otherwise holds its maximum value. The computation time for this example was 0.86~sec using the structure exploiting MIQP solver, and 2.75~sec using Gurobi with H-rep polytopes.

\subsection{Case Study 2: Package Delivery Drone with Terminal SOC Constraints}

\begin{figure}
    \centering
    \begin{subfigure}[b]{0.49\linewidth}
        \centering
        \adjustbox{trim=0in 0.1in 0in 0.1in}{%
            \includegraphics[width=\linewidth]{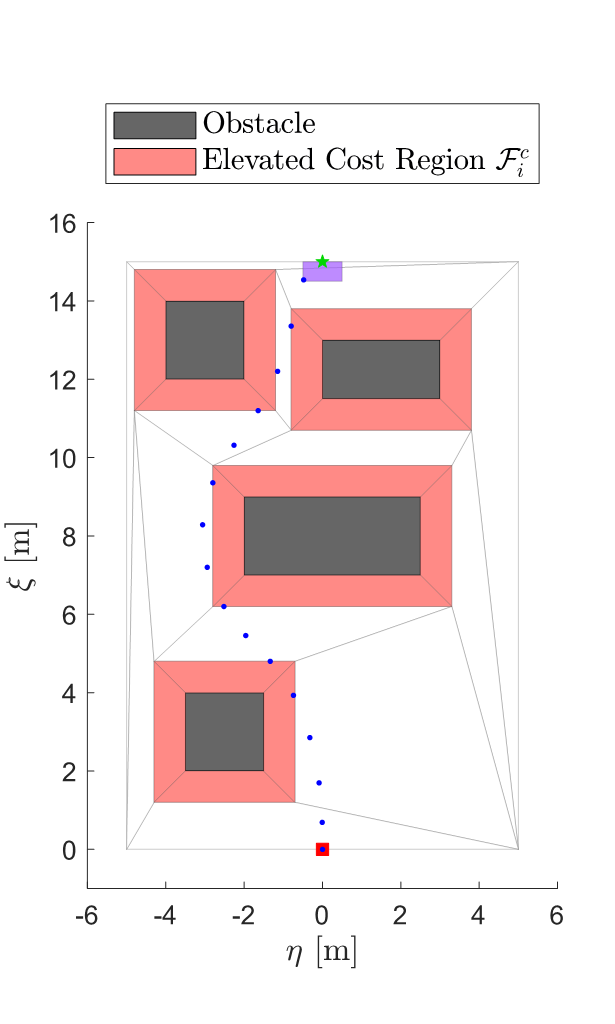}
        }
        \caption{$SOC_N \in [0.9, 1.0]$}
        \label{fig:ex2-traj}
    \end{subfigure}
    \begin{subfigure}[b]{0.49\linewidth}
        \centering
        \adjustbox{trim=0in 0.1in 0in 0.1in}{%
            \includegraphics[width=\linewidth]{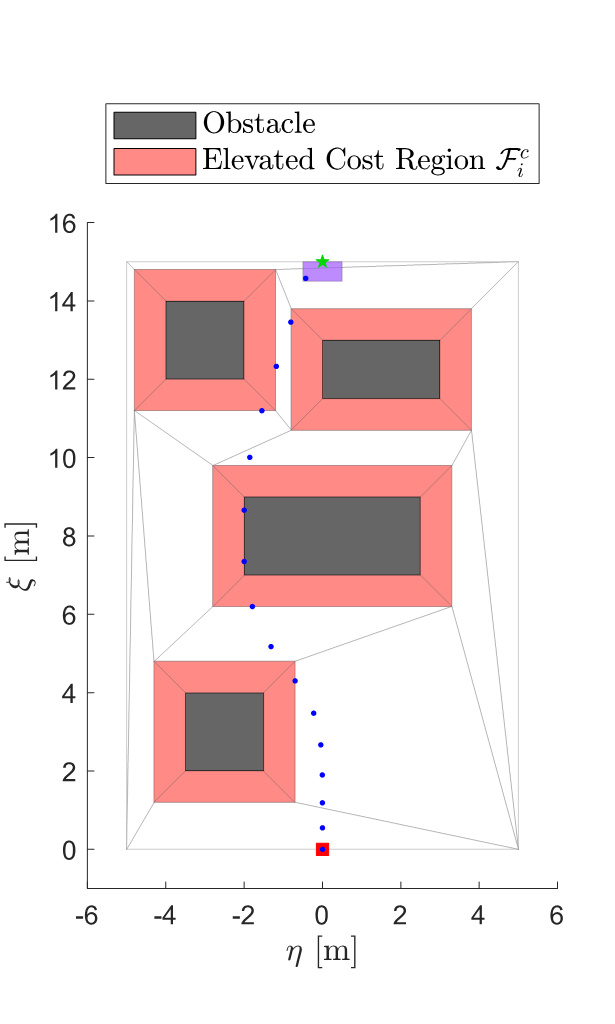}
        }
        \caption{$SOC_N \in [0.93, 1.0]$}
        \label{fig:ex3-traj}
    \end{subfigure}
    \caption{\emph{Case Study 2:} Planned trajectory for package delivery drone example given two different terminal state of charge constraints.}
\end{figure}

\begin{figure*}
    \centering
    \begin{subfigure}[t]{0.49\linewidth}
        \centering
        \includegraphics[width=\linewidth]{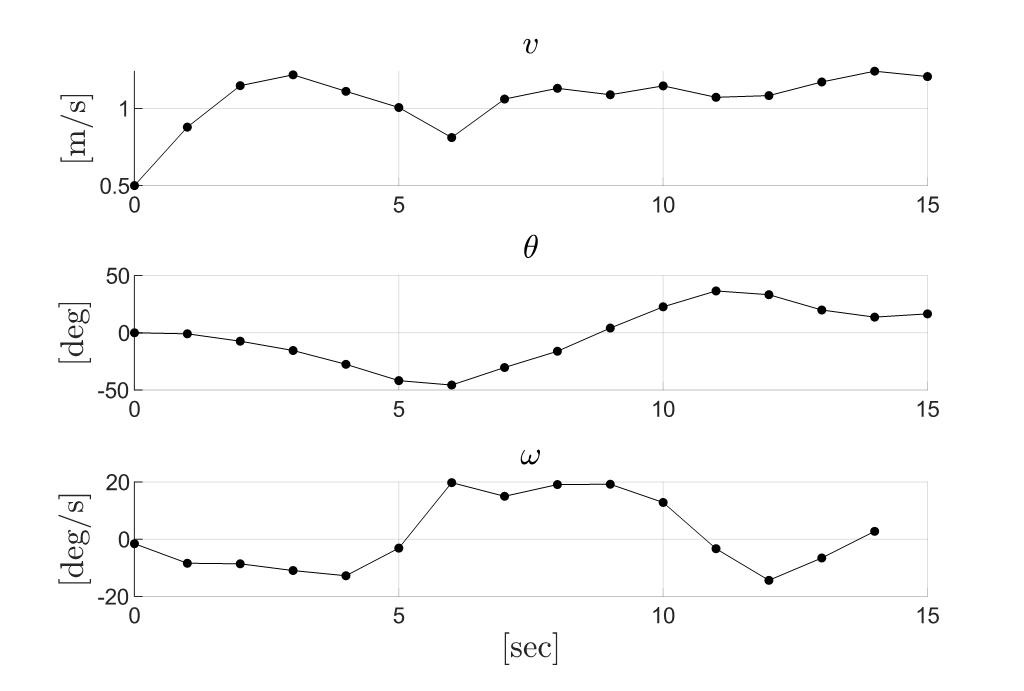}
        \caption{Motion states and inputs, $SOC_N \in [0.9,1.0]$}
        \label{fig:ex2-motion-states}
    \end{subfigure}
    \begin{subfigure}[t]{0.49\linewidth}
        \centering
        \includegraphics[width=\linewidth]{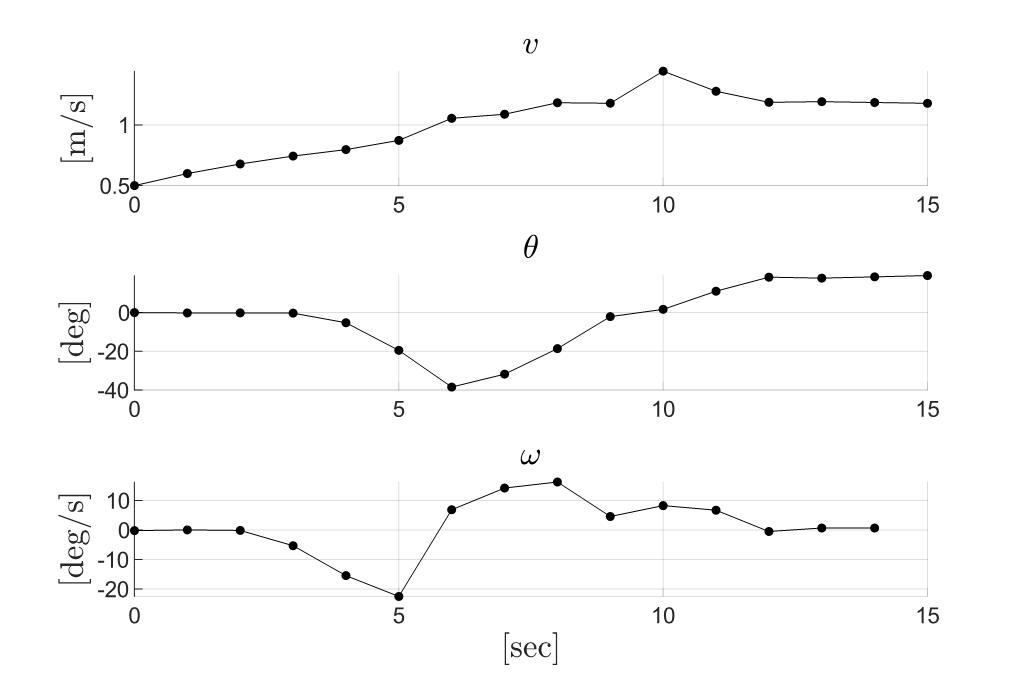}
        \caption{Motion states and inputs, $SOC_N \in [0.93,1.0]$}
        \label{fig:ex3-motion-states}
    \end{subfigure}
    \begin{subfigure}[t]{0.49\linewidth}
        \centering
        \includegraphics[width=\linewidth]{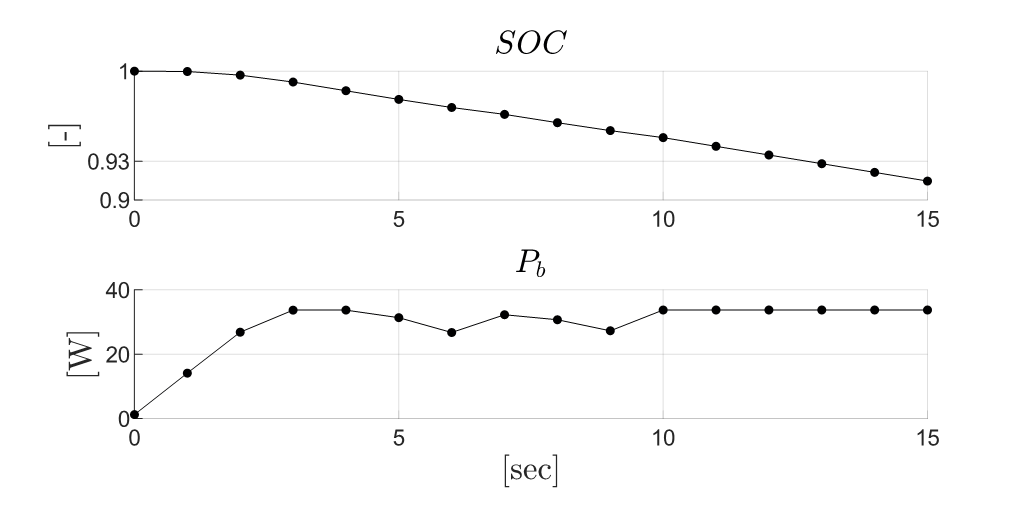}
        \caption{Energy states, $SOC_N \in [0.9,1.0]$}
        \label{fig:ex2-energy-states}
    \end{subfigure}
    \begin{subfigure}[t]{0.49\linewidth}
        \centering
        \includegraphics[width=\linewidth]{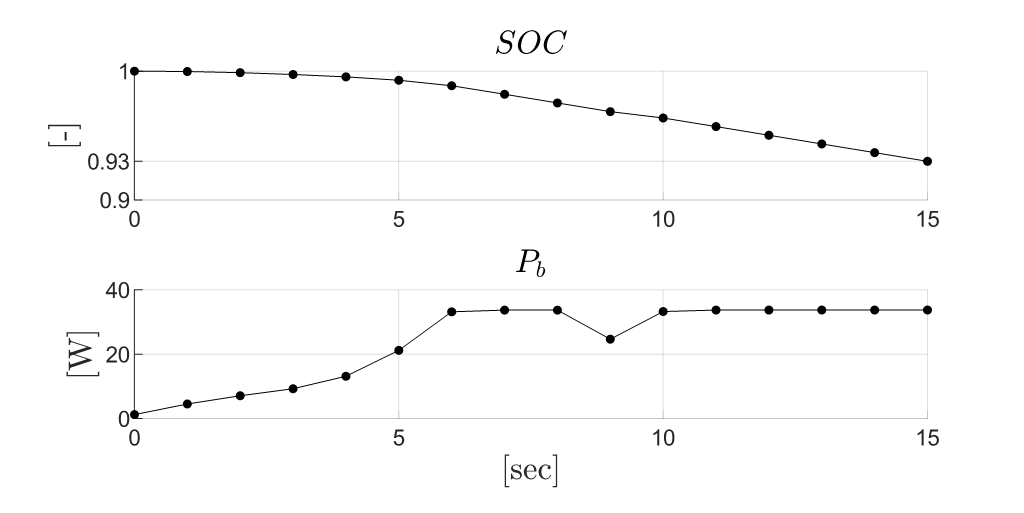}
        \caption{Energy states, $SOC_N \in [0.93,1.0]$}
        \label{fig:ex3-energy-states}
    \end{subfigure}
    \caption{\emph{Case Study 2:} Planned energy and motion states / inputs for the package delivery drone example given two different terminal state of charge constraints.}

    \vspace{-0.2in}
    
\end{figure*}

This example considers an electrically-powered package delivery drone. The MPC motion and energy utilization planner must find a system trajectory to a wayset that is provided by a higher level mission planner. The wayset contains both position constraints and a battery state of charge constraint that must be achieved in order to ensure the vehicle can complete its mission. See \cite{shekhar2015robust, koeln2020vertical} for references on waysets. The generated motion plan must avoid obstacles (e.g., buildings) in the environment. To discourage flying close to obstacles, neighboring regions of the obstacle-free space are assigned a fixed cost as described in Sec.~\ref{sec:mpc-formulation}. 

The wayset constraints are added to the terminal constraint set $\mathcal{X}_N$ by modifying \eqref{eq:state-conzono} as
\begin{multline}
    \mathcal{Z}_{cxN} = \left\langle \text{blkdiag} \left( \begin{bmatrix}
        G_{cx} & g_{\xi N} & g_{\eta N} & g_{SOC N}
    \end{bmatrix}\right), \right. \\ 
    \left. \begin{bmatrix} 
        \mathbf{c}_{cx}^T & c_{\xi N} & c_{\eta N} & c_{SOC N} 
    \end{bmatrix}^T,  \begin{bmatrix}
        A_{cx} & 0 & 0 & 0 
    \end{bmatrix}, \mathbf{b}_{cx} \right\rangle \;,
\end{multline}
where $\mathcal{Z}_{cx} = \left\langle G_{cx}, \mathbf{c}_{cx}, A_{cx}, \mathbf{b}_{cx} \right\rangle$. The parameters $g_{\xi N}$, $g_{\eta N}$, $g_{SOC N}$, $c_{\xi N}$, $c_{\eta N}$, and $c_{SOC N}$ are defined according to the maximum and minimum values of $\xi$, $\eta$, and $SOC$ within the wayset as in \eqref{eq:state-conzono-vars}.

Maximum and minimum states and inputs for this model are given in Table~\ref{tab:ex23-state-limits}. There is no engine for the vehicle in this example, so $m_f$, $P_e$ and $\dot{P}_e$ are eliminated from the equations of motion and constraints. The battery capacity is set to $C_b = 5000~J$. The velocity minimum $v_{min}$ is only used in the turn rate limit constraint \eqref{eq:motion-constr-omega} and does not enter the state constraints in contrast with \emph{Case Study 1}. A discretization of the continuous-time dynamics model from Sec.~\ref{sec:reduced-order-model} is used with a discrete time step of $\Delta t = 1$~sec.

\begin{table}[htbp]
  \setlength{\tabcolsep}{3pt}
    \centering
        \caption{\emph{Case Study 2:} State and input limits for package delivery drone model}
    \begin{tabular}{c|c c c c c}
         \textbf{Parameter} & $v$ & $\omega$ & $SOC$ & $P_b$ & $\dot{P}_b$ \\ \toprule
         \textbf{units} & [m/s] & [deg/s] & [] & [W]  & [W/s] \\ \midrule
         \textbf{min} & 0.5 & -45 & 0.25 & 1.25 & -20.0 \\
         \textbf{max} & 1.5 & 45 & 1 & 33.75 & 20.0
    \end{tabular}
    \label{tab:ex23-state-limits}
\end{table}

The cost function parameters are given as
\begin{subequations}
\begin{align}
    &Q_k = 0_{6 \times 6} \;, \\
    &Q_N = \text{diag}([10, 0, 10, 0, 0, 0]) \;, \\
    &R_k = \text{diag}([1, 1, 10^{-4}]) \;, \\
    &\mathbf{q}_k = \mathbf{q}_N = [0, 0, 0, 0, 0, 10^{-3}]^T \;, \\
    &q^r(\mathbf{x}_k) = \begin{cases}
        10 \;,\; \begin{bmatrix} \xi_k & \eta_k \end{bmatrix}^T \in \mathcal{F}^c \\
        0 \;,\; \text{otherwise}
    \end{cases} \;,
\end{align}
\end{subequations}
where $\mathcal{F}^c = \bigcup_i \mathcal{F}^c_i$ denotes the union of elevated-cost regions of the map.

For the case that the wayset has a battery state of charge constraint of $SOC_N \in [0.9, 1.0]$, the optimal motion and energy utilization plan is given in Figs.~\ref{fig:ex2-traj}, \ref{fig:ex2-motion-states}, and \ref{fig:ex2-energy-states}. Here, the vehicle avoids the elevated cost regions and maintains some separation from the obstacles as desired. Changing the state of charge constraint to $SOC_N \in [0.93, 1.0]$ results in the optimal motion and energy utilization plan given in Figs.~\ref{fig:ex3-traj}, \ref{fig:ex3-motion-states}, and \ref{fig:ex3-energy-states}. In this case, the vehicle must pass close to an obstacle through one of the elevated cost regions in order to satisfy the terminal state of charge constraint. Using the structure-exploiting MIQP solver, the MPC solution times were 0.96~sec and 0.83~sec for $SOC_N \in [0.9,1.0]$ and $SOC_N \in [0.93, 1.0]$, respectively. Using Gurobi with H-rep polytopes, the corresponding solution times were 2.46~sec and 3.93~sec.

\section{Conclusions} \label{sec:conclusion}

A mixed-integer MPC formulation for coupled motion and energy utilization planning of a UAS was presented. 
The resulting energy-aware motion planner jointly considers motion and energy specifications on the planned trajectory. Specifications presented in this paper include obstacle avoidance, engine power restrictions when flying over noise-restricted areas, and terminal battery state of charge requirements. By leveraging constrained zonotope and hybrid zonotope constraint representations within a mixed-integer quadratic program solver designed to exploit the structures of those representations, MPC optimization times of less than 1~sec are achieved, indicating that the proposed approach is tractable for real-time implementation.

Future work will include developing a low-level controller to track the motion and energy plans generated using this MPC controller. A high-level planner will also be developed that provides references and/or waysets to the energy-aware motion planner. A non-convex state constraint set will also be considered in order to allow for more flexible planning subject to a velocity minimum as described in Sec.~\ref{sec:reduced-order-model}. The resulting multi-layer control architecture will be evaluated in simulation using higher fidelity nonlinear dynamics models for the aircraft motion dynamics and energy systems.

\bibliography{bibitems}
\bibliographystyle{ieeetr}

\end{document}